\documentclass[doublecol]{epl2} 
\usepackage{here}
\usepackage{amsmath}
\usepackage{array}
\usepackage{amssymb}
\usepackage[dvipdfmx]{hyperref}

\usepackage{lipsum}

\newcommand{\dv}[2]{\frac{\mathrm{d}{#1}}{\mathrm{d}{#2}}}
\newcommand{\leri}[1]{\left({#1}\right)}

\usepackage{color}
\usepackage{ulem}

\usepackage{xcolor}

\title{Effect of gravitational field on collective motion of fish}
\shorttitle{Effect of gravitational field on collective motion of fish} 

\author{S. Ito \and N. Uchida \thanks{E-mail: \email{uchida@cmpt.phys.tohoku.ac.jp}}}
\shortauthor{S. Ito and N. Uchida}

\institute{                    
 Tohoku University, Department of Physics, Sendai, 980-8578, Japan
}

\abstract{
Fish exhibit various patterns of collective motion, in which individual 
fish sense the gravitational field and tend to move horizonally.
We study the effect of gravity on the collective patterns
by incorporating suppression of vertical motion
in an agent-based model.
The gravitational factor induces a tornado
which is a vertically and highly elongated form of a torus,
The vortex axis becomes almost identical to the vertical axis
even when the gravitational factor is weak compared to the interaction between fish.
We also obtained a vertically elongated polarized school with high frontal density.
Our results clarify
the effect of gravity on the shape of clusters, 
individual-level motion, and mobility of the entire cluster.
}

\begin{document}

\maketitle


\section{Introduction}
Collective motion is ubiquitously found in life of various organisms
such as birds, fish, insects and mammals~\cite{Vicsek2012}.
Patterns of highly coherent motion include polarized schools 
and rotating clusters.
Fish exhibit a wide variety of rotating clusters such as 
tori, rings, balls, and tornadoes \cite{Parrish2002,Lopez2012}.
They are especially interesting and non-trivial because the rotational symmetry is spontaneously broken 
by interaction between individual fish \cite{Tunstrom2013}.
Theoretically, agent-based models \cite{Couzin2002,DOrsogna2006,Nguyen2012,Calovi2014,Chuang2016,Filella2018} 
have been proposed and reproduced a torus and a ring.
In addition, a ball-shaped rotating cluster (resembling a ``bait-ball"~\cite{Lopez2012}) 
is reproduced in a recent model by the present authors~\cite{Ito2021}.
The model uses 
topological interaction \cite{Ballerini2007}and some experimentally observed behaviors of fish,
which facilitate
formation of a giant rotating cluster.

In Nature, 
the vortex axis of a rotating cluster of fish is oriented almost vertically \cite{Terayama2015}, presumably due to gravity.
Actually, the otolith (``the ear stone") allows the fish to perceive gravity
and swim almost perpendicular to the water surface \cite{Anken2010,Mirbach2019}.
In the previous models \cite{Couzin2002,Nguyen2012,Chuang2016,Ito2021}, however,
the direction of the vortex axis is not constrained in any particular direction.
A systematic study of the effect of 
gravity on collective motion of fish has been lacking,
although a previous study~\cite{Hemelrijk2008}
introduced it as a fixed parameter 
and yielded a polarized school with high frontal density.
Moreover, 
there are no previous models showing 
a tornado which is a vertically and highly elongated form of a torus~\cite{Parrish2002}.
We consider that constraints on the vertical motion of fish is
important for the occurrence of a tornado.

In this Letter, we introduce a gravitational factor into our previous model~\cite{Ito2021}
and study its effect on collective motion.
We find emergence of
a giant tornado cluster and a polarized school flattened along the direction of movement.
We also study the effect of gravity on the individual-level motion
and the mobility of the entire cluster.

\begin{figure*}[!t]
\centering
\includegraphics[width=\linewidth,bb=0 0 360 252]{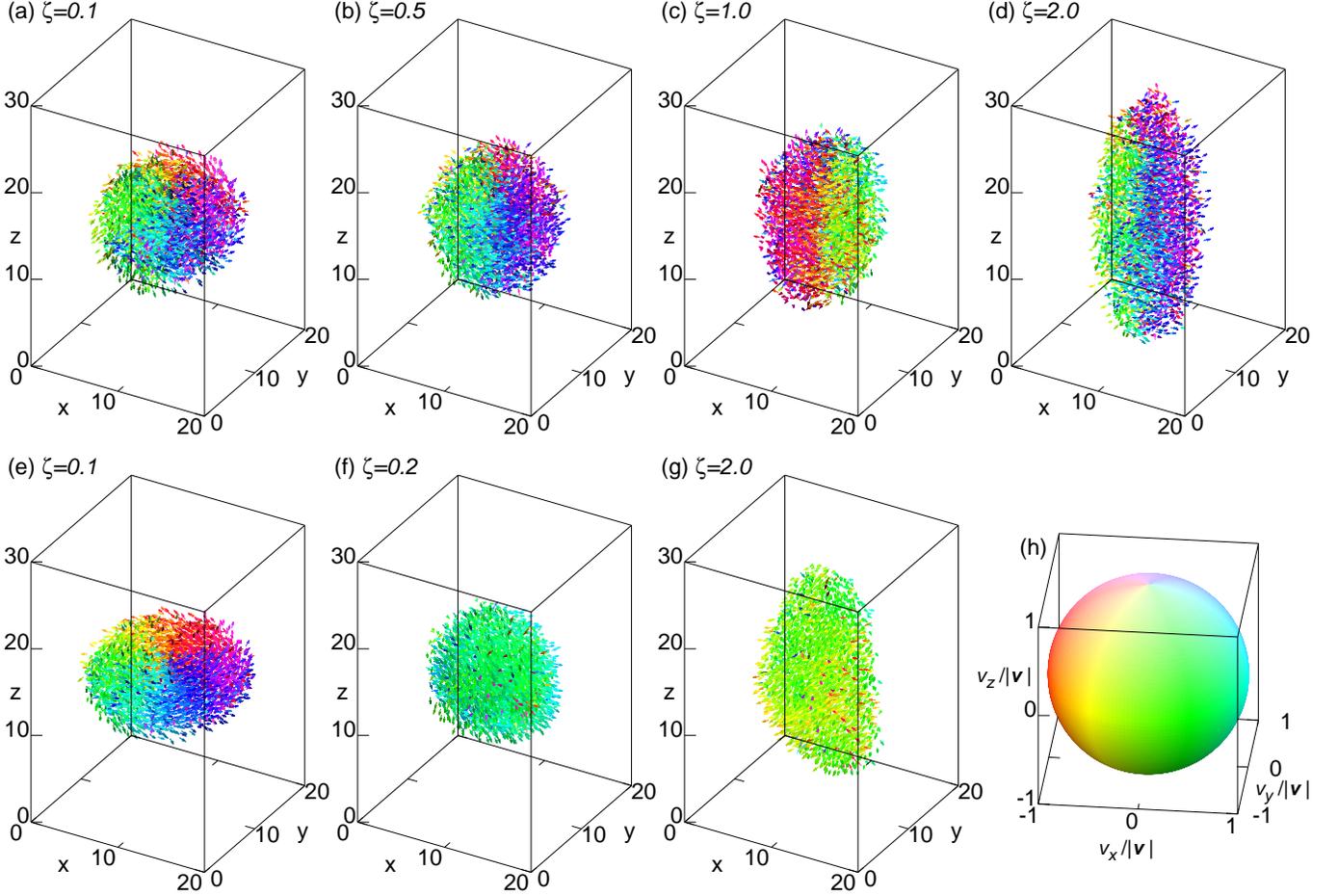}
\caption{Snapshots of clusters at the end of simulations ($t=3000$).
Agents are represented by arrows of length 2/3 (=1 BL), and the color corresponds to the moving direction of the agent $\bm{v}/|\bm{v}|=(v_x/|\bm{v}|,v_y/|\bm{v}|,v_z/|\bm{v}|)$ according to the color sphere (h).
Shown are only a part of the simulation box that contains the agents and the origin is shifted for visibility. 
(a-d) rotating clusters for 
$N_u=1,\lambda=9.0$ with $\zeta=0.1,0.5,1.0,2.0$.
(e-g) a rotating cluster and polarized schools for
$N_u=3,\lambda=9.0$ with $\zeta=0.1,0.2,2.0$.
}
\label{snapshot}
\end{figure*}

\section{Model}

We introduce the equation of motion of fish as

\begin{eqnarray}
\label{velocity}
\tau_0\dv{\bm{v}_i}{t}&=&(v_0-|\bm{v}_i|)\frac{\bm{v}_i}{|\bm{v}_i|}+\frac{1}{|\mathcal{L}_{i}|}\sum_{j \in \mathcal{L}_{i}}g(|\bm{r}_{ij}|)(\bm{v}_j-\bm{v}_i)\nonumber\\
&&+\frac{1}{|\mathcal{L}_{i}|}\sum_{j \in \mathcal{L}_{i}}g(|\bm{r}_{ij}|)\left(v_r\frac{\bm{r}_{ij}}{|\bm{r}_{ij}|}-\bm{v}_i\right)\nonumber\\
&&+\frac{\Lambda_i(t)}{|\mathcal{A}_{i}|}\sum_{j \in \mathcal{A}_{i}}\left(v_a\frac{\bm{r}_{ji}}{|\bm{r}_{ji}|}-\bm{v}_i\right)-\zeta v_{z,i}\bm{e}_z,
\end{eqnarray}
where $\bm{r}_i$ and $\bm{v}_i=d\bm{r}_i/dt$ ($i=1,2,\ldots, N$) is the position and velocity of the $i$-th agent, respectively, with 
$\bm{r}_{ij}=\bm{r}_i-\bm{r}_j$, 
$v_{z,i}$ being 
the $z$-component of $\bm{v}_i$, and $\bm{e}_z$ 
is the unit vector along the vertical axis.
The speed is set to relax to
$v_0$ for an isolated agent, 
and the $z$-component of the velocity 
relaxes to zero with the time-scale $\tau_0/\zeta$: 
the gravitational factor $\zeta$ 
describes the sensitivity of fish to
the gravitational field~\cite{Hemelrijk2008}.

As for the interactions, there are reorientation, repulsion, and attraction 
with the neighbor agents, which are expressed by
the second, third, and 
fourth term on the right-hand side of Eq. (\ref{velocity}), respectively.
Firstly, the orientational and repulsive interactions act with up to $N_u$-th nearest neighbors in the radius $r_e$.
$N_u$ is called the interaction capacity.
The set of agents that can interact with the $i$-th agent by these
interactions is denoted by $\mathcal{L}_i$, and the number of them by $|\mathcal{L}_i|$.
$v_r$ is the speed of collision avoidance, and the function
\begin{align}
\label{exclusion func}
g(|\bm{r}|)= \left\{ \begin{array}{ll}
\frac{r_b}{|\bm{r}|} & [|\bm{r}|\leq r_b], \\
1 & [|\bm{r}|>r_b]
\end{array} \right.
\end{align}
expresses the excluded volume effect (where $r_b$ is the body length).

The attractive interaction operates at distances between $r_e$ and $r_a$, and the set of agents that can attract the $i$-th agent is denoted by $\mathcal{A}_{i}=\left\{j\middle|r_e<|\bm{r}_{ij}|\leq r_a\right\}$
with its size $|\mathcal{A}_i|$.
$v_a$ is the speed of fast-start, which an acceleration mode of fish to escape from predators.
The time-dependent strength of attraction $\Lambda_i(t)$
is introduced to model the screening of attraction in a cluster \cite{Katz2011} 
and the duration of attraction by fast-start \cite{Domenici1997}.
The attraction is turned on ($\Lambda_i(t)=\lambda$) at the moment when $|\mathcal{L}_i(t)|$ becomes smaller than $N_u$, and lasts for the duration $\tau$. If $|\mathcal{L}_i(t)|=N_u$ after the period $\tau$, the attraction is switched off ($\Lambda_i(t)=0$). Otherwise, the attraction is maintained for another period $\tau$, and this will be repeated until we finally get $|\mathcal{L}_i(t)|=N_u$.

In our simulation, all
lengths and time are rescaled by the radius of equilibrium $r_e=1.5$ BL (body length) and the characteristic timescale $\tau_0=1$ sec, respectively.
We fixed the parameter values
$N=3000,~r_e=1,~\tau_0=1,~r_b=2/3,~r_a=5,~v_0=1,~v_r=1,~v_a=5,~\tau=0.1$,
and regarded
$N_u,~\lambda,~\zeta$ as the control parameters.
We focus on
the parameter region $1 \le N_u \le 3$ and
$5.0 \le \lambda \le 9.0$ where rotating clusters
of various shapes emerge 
in the absence of gravity~\cite{Ito2021}.
(See Ref.~\cite{Ito2021} 
for further details of the model 
and the choice of parameter values.)

We used
the Runge-Kutta method for numerical integration of the equation of motion 
with the time step $dt=0.005$.
The simulation box is a cuboid whose 
dimensions are $L=40$
in the $x$ and $y$ directions
and $L_z=50$
in the $z$ direction 
with the periodic boundary condition.
Unless otherwise stated, we use the initial condition in which the agents are randomly distributed in a single spherical cluster aligned in
the same ($x$) direction and speed $v_0$. 
The radius of the sphere was chosen for each set of $(N_u, \lambda)$ 
so that a single cluster is obtained and maintained for $\zeta=0$. 
The same initial radius was used also for $\zeta>0$, 
for which a single-cluster state is maintained except for rare cases.
This initial condition helps reducing the computational cost,
while it was checked that the final (dynamical steady) states
do not depend on the initial conditions~\cite{Ito2021}.


\section{Results: cluster shapes}

Fig.~\ref{snapshot} shows typical snapshots at 
the end of the simulations ($t=3000$) when the clusters reach dynamically steady states.
Fig.~\ref{snapshot} (a-d) 
demonstrate that rotating clusters are
obtained for $N_u=1$ and become
elongated as $\zeta$ increases; a ball-shaped cluster changes to a vertically elongated cylinder, which we call a tornado.
On the other hand, when $N_u=3$, a torus-shaped rotating cluster is transformed into a polarized school in
which the agents have almost the same direction, 
and then the cluster is elongated as $\zeta$ increases.
The time evolution of the clusters are shown in
Movie 1 ($N_u=1,\lambda=9.0,\zeta=2.0$) and Movie 2 ($N_u=3,\lambda=9.0,\zeta=2.0$) 
in the Supplementary Material.
We checked that the same patterns
emerge with the initial condition in which the agents 
have randomly distributed positions and directions: see Movies 3, 4.

\begin{figure}[!t]
\centering
\includegraphics[width=\linewidth,bb=0 0 360 252]{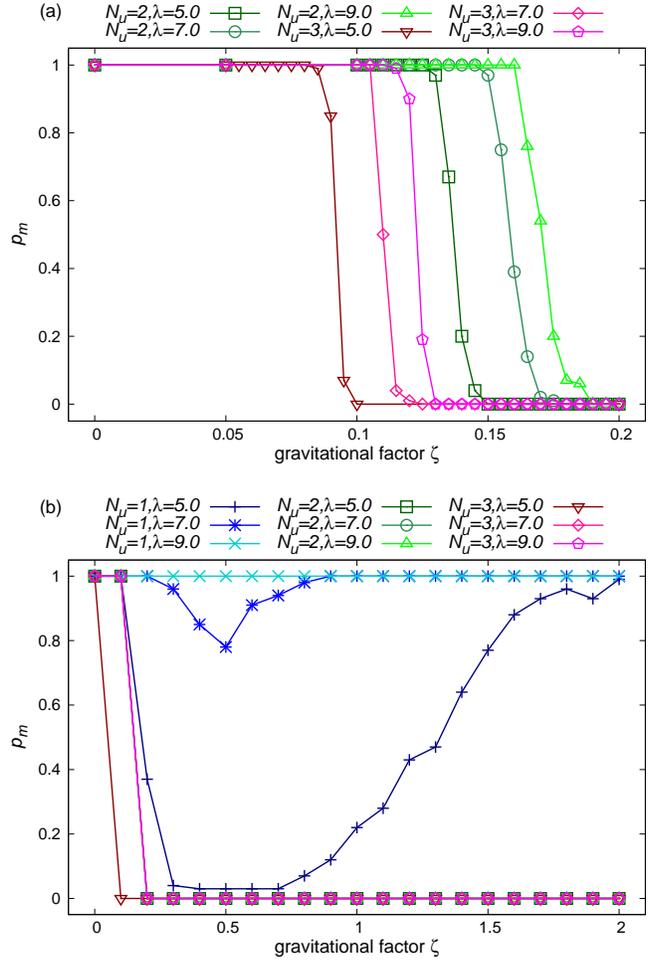}
\caption{
The probability of emergence of rotating clusters $p_m$ 
as a function of $\zeta$ in the case of (a) a narrow range $\zeta\in[0.0,0.2]$ and (b) a wide range $\zeta\in[0.0,2.0]$.
}
\label{pmill}
\end{figure}

In order to distinguish between a rotating cluster and a polarized school, 
we use 
the polar order parameter $\bm{P}$
and the rotational order parameter $\bm{M}$. They are defined by
\begin{equation}
\label{order}
\bm{P}(t) =\frac{1}{N}\sum_{i=1}^N\frac{\bm{v}_i(t)}{|\bm{v}_i(t)|},~\bm{M}(t)=\frac{1}{N}\sum_{i=1}^N\frac{\bm{c}_i(t)\times\bm{v}_i(t)}{|\bm{c}_i(t)||\bm{v}_i(t)|},
\end{equation}
where $\bm{c}_i=\bm{r}_i-\bm{r}_G$ with $\bm{r}_G$ being the center of mass position of the cluster.
The magnitude of the polar order parameter $P(t)$ 
becomes unity for a completely aligned 
polarized school, and 
is (almost) zero for random swarming states and 
rotating clusters with a cylindrical symmetry.
On the other hand,
the magnitude of the rotational order parameter $M(t)$ 
becomes unity for
a ring-shaped rotating cluster, and is zero 
for a completely ordered polarized school with central symmetry to center of mass.
For a rotating cylinder, $M$ is a decreasing function of the aspect ratio,
and is calculated in Appendix A.
We define a rotating cluster by the criterion $\overline{M}>\overline{P}$.
Here and hereafter, $\overline{X}$ is 
the time-average of any quantity $X(t)$ over the time interval $[2500,3000]$.
The probability of
emergence of a rotating cluster 
$p_m$ is defined by the number of cases where
$\overline{M}>\overline{P}$ 
divided by the number of samples (n$=100$)
in which a single cluster
is maintained until the end of a simulation ($t=3000$).
We plot $p_m$ as a function of $\zeta$ in Fig.~\ref{pmill}.
When $N_u=$ 2 or 3, 
$p_m$ decreases 
around $\zeta \sim 0.1$
(Fig.~\ref{pmill}(a)), 
and becomes zero for larger $\zeta$
(i.e. the cluster is a polarized school) 
as shown in Fig.~\ref{pmill}(b).
Dominance of rotating clusters ($p_m=1$)
is maintained up to larger values of $\zeta$ 
as $N_u$ is smaller or $\lambda$ is larger.
On the other hand, when $N_u=1$, 
$p_m$ tends to decrease 
once, but then increases again and reaches 1 
as $\zeta$ is increased
(Fig.~\ref{pmill}(b)).
For $\lambda=9.0$, $p_m=1$ is maintained 
for any $\zeta$.

\begin{figure}[!t]
\centering
\includegraphics[width=\linewidth,bb=0 0 360 252]{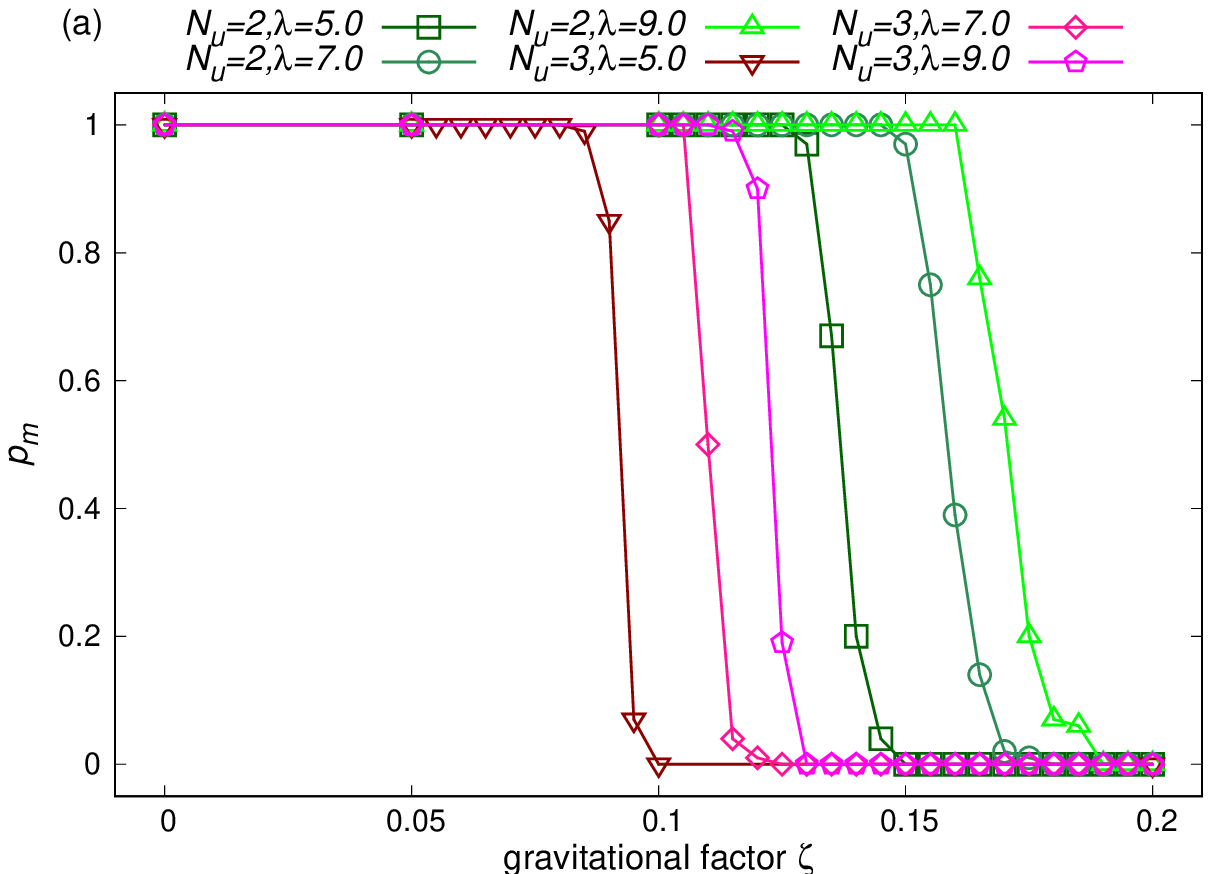}
\end{figure}

Next, we define some quantities 
that characterize the size and shape of the clusters.
The outer radius of the cluster 
is defined as
\begin{equation}
\label{Router}
R_o(t)=\max_{i=1,\cdots,N}\left|\bm{c}_i^\perp(t)\right|,
\end{equation}
where the radial position $\bm{c}_i^\perp$ 
is the projection of $\bm{c}_i$ onto a vector which is perpendicular to $\bm{M}(t)$ and $\bm{c}_i \times\bm{M}(t)
$.
The cluster shape is characterized by
the moment of inertia tensor 
$I_{\mu\nu}(t)=\sum_{i=1}^N\left[\bm{c}_i^2(t)\delta_{\mu\nu}-c_{i,\mu}(t)c_{i,\nu}(t)\right],$
where $\mu, \nu = x,y,z$ and $\delta_{\mu\nu}$ is Kronecker delta.
Diagonalizing it and using its eigenvectors $\bm{u}_\alpha(t)$ ($\alpha=1,2,3$),
with the principal moments of inertia satisfying $I_1(t)\geq I_2(t)\geq I_3(t)$,
we define the cluster sizes by
\begin{equation}
\label{Salpha}
S_\alpha(t)=\max_{i=1,\cdots,N}\left|\bm{c}_i(t)\cdot\bm{u}_\alpha(t)\right|,
\end{equation}
where $\alpha=1,2,3$.
Note that $S_3$ corresponds to the semi-major axis 
for rod-shaped cluster and 
to the outer radius for a torus-shaped cluster.

\begin{figure}[!t]
\centering
\includegraphics[width=\linewidth,bb=0 0 360 252]{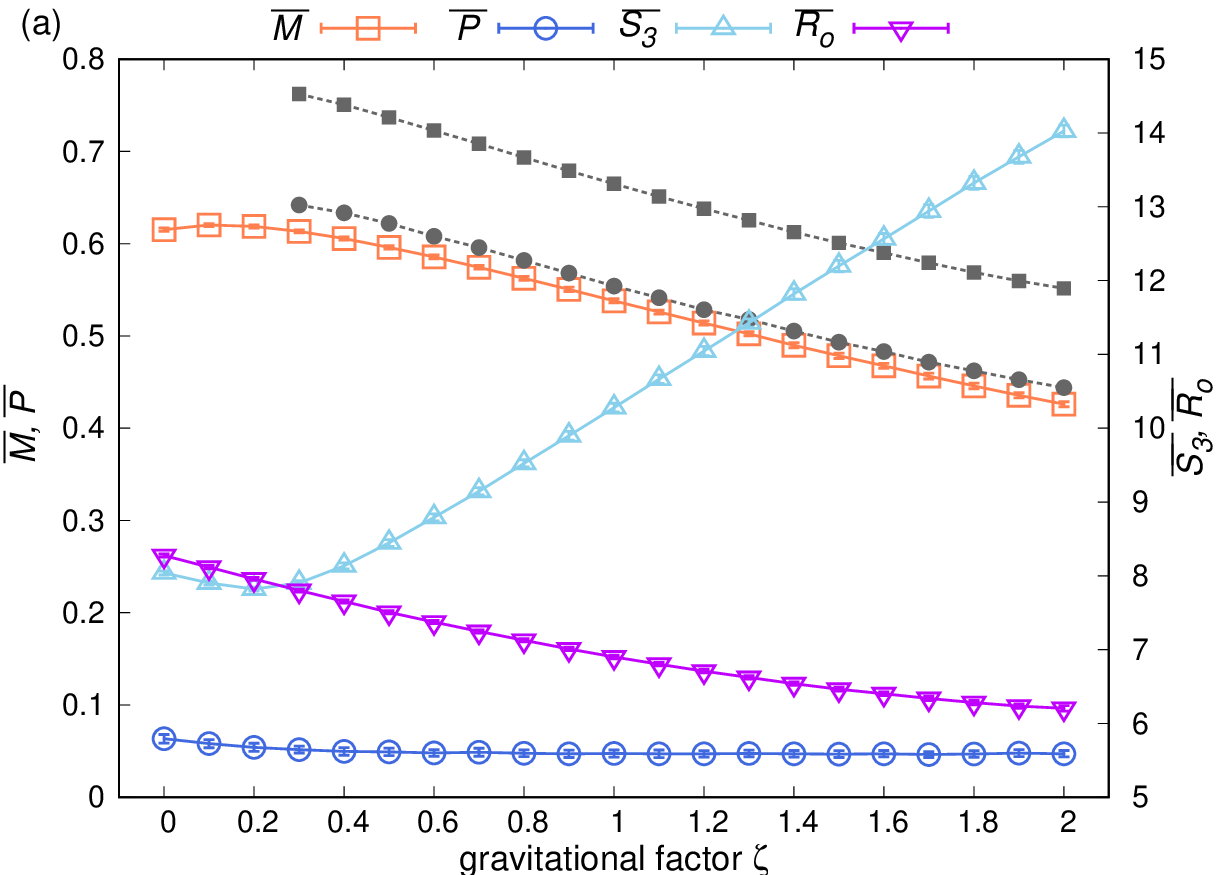}
\end{figure}
\begin{figure}[!t]
\centering
\includegraphics[width=\linewidth,bb=0 0 360 252]{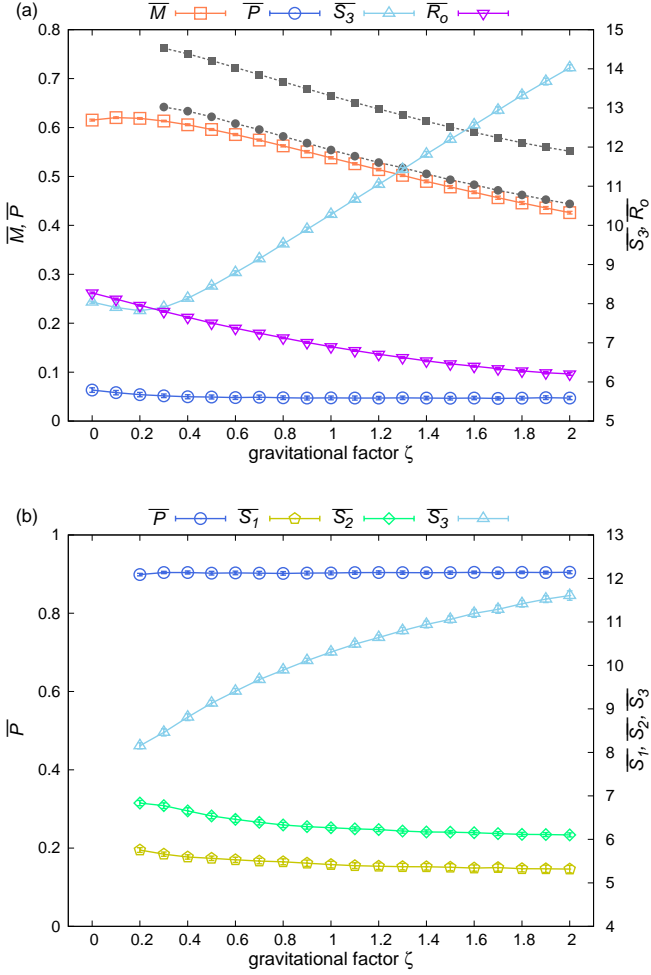}
\caption{The time-averaged order parameters $\overline{P},\overline{M}$ and the time-averaged cluster sizes $\overline{R_o},\overline{S_1},\overline{S_2},\overline{S_3}$ as functions of $\zeta$, for
(a) $N_u=1,\lambda=9.0$ and (b) $N_u=3,\lambda=9.0$.
An ensemble average over 100 simulations is taken for each point,
and the error bars represent the standard deviation.
In (a), the dashed lines with the squares and circles show
the calculated values for a cylinder,
$M=\Psi(\overline{S_3}/\overline{R_o})$
and
$M=w(v_r,v_\phi)\Psi(\overline{S_3}/\overline{R_o})$, respectively. 
}
\label{order size}
\end{figure}

In Fig.~\ref{order size}, we plot the time-averaged order parameters $\overline{P}$, $\overline{M}$
as well as the cluster sizes $\overline{R_o}$, $\overline{S_1}$, $\overline{S_2}$, $\overline{S_3}$
as functions of $\zeta$.
Fig.~\ref{order size}(a) shows that, for $N_u=1,\lambda=9.0$
(which correspond to Fig.~\ref{snapshot}(a-d)), 
$\overline{R_o}$ monotonically decreases
while $\overline{S_3}$ 
starts to increase at $\zeta \sim 0.3$ and is deviated from $\overline{R_o}$.
The height of the cluster 
($=2\overline{S_3}$) 
reaches 28 at the maximum,
which is much larger than the interaction range $r_a=5$.
The order parameter $\overline{P}$ 
takes very small values ($<0.1$) for any $\zeta$,
while $\overline{M}$ 
starts at a larger value $\sim 0.6$ and
decreases from $\zeta\sim 0.3$.
The decrease in $\overline{M}$ 
is well described by the calculated value
$M=w(v_r,v_\phi)\Psi(\overline{S_3}/\overline{R_o})$ 
for a cylinder-shaped rotating cluster.
(See Appendix A for the definitions of $w$ and $\Psi$.)
These results quantitatively show
that the cluster has a giant tornado structure 
for $N_u=1,\lambda=9.0$, and $\zeta =2.0$. 

Fig.~\ref{order size}(b) shows that, 
when $N_u=3,\lambda=9.0$ 
(which correspond to polarized schools in Fig.~\ref{snapshot}(f-g)), 
$\overline{P}$ keeps a large constant value $\sim 0.9$.  
The aspect ratio $\overline{S_3}/\overline{S_2}$ increases sharply with $\zeta$
and is close to two at $\zeta=2$. 
The inequality $\overline{S_3} > \overline{S_2} > \overline{S_1}$ holds
for any $\zeta$ showing that the cluster shape is biaxially distorted.

\begin{figure}[!t]
\centering
\includegraphics[width=\linewidth,bb=0 0 360 252]{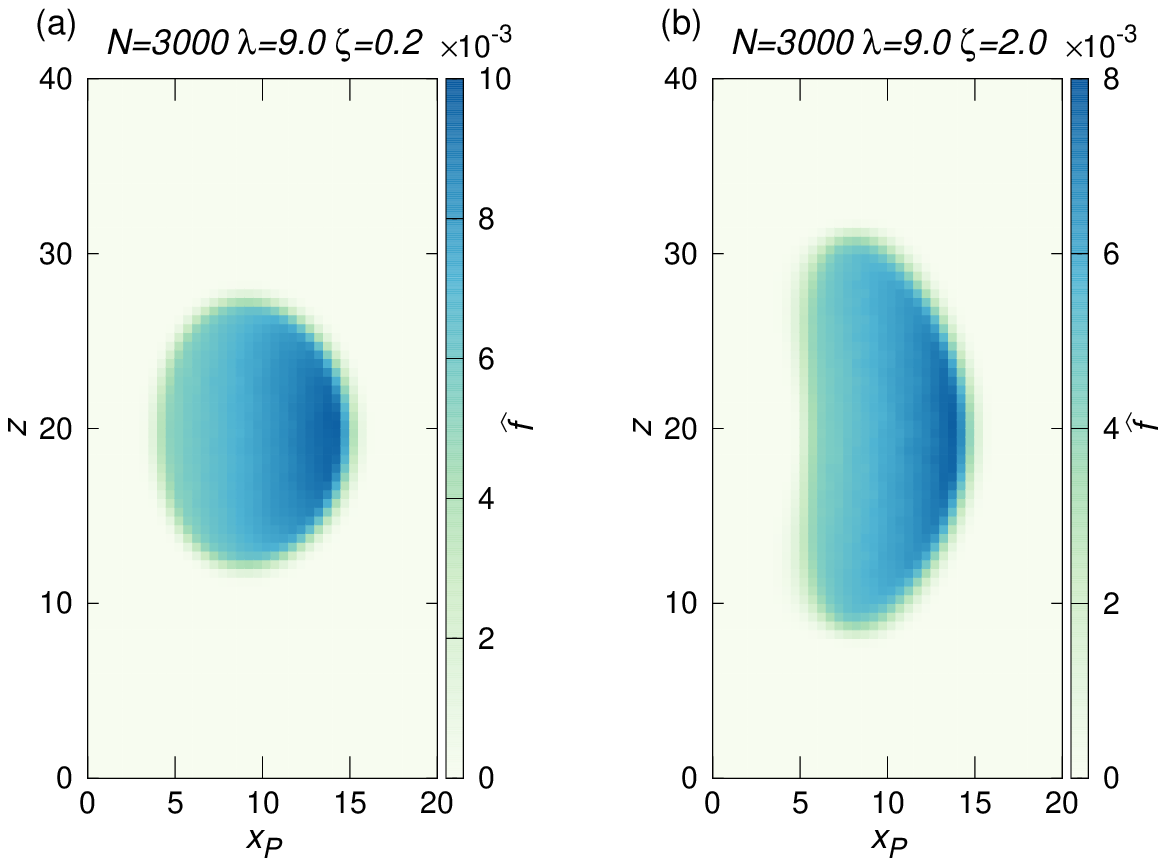}
\vspace{-10mm}\\
\includegraphics[width=\linewidth,bb=0 0 360 252]{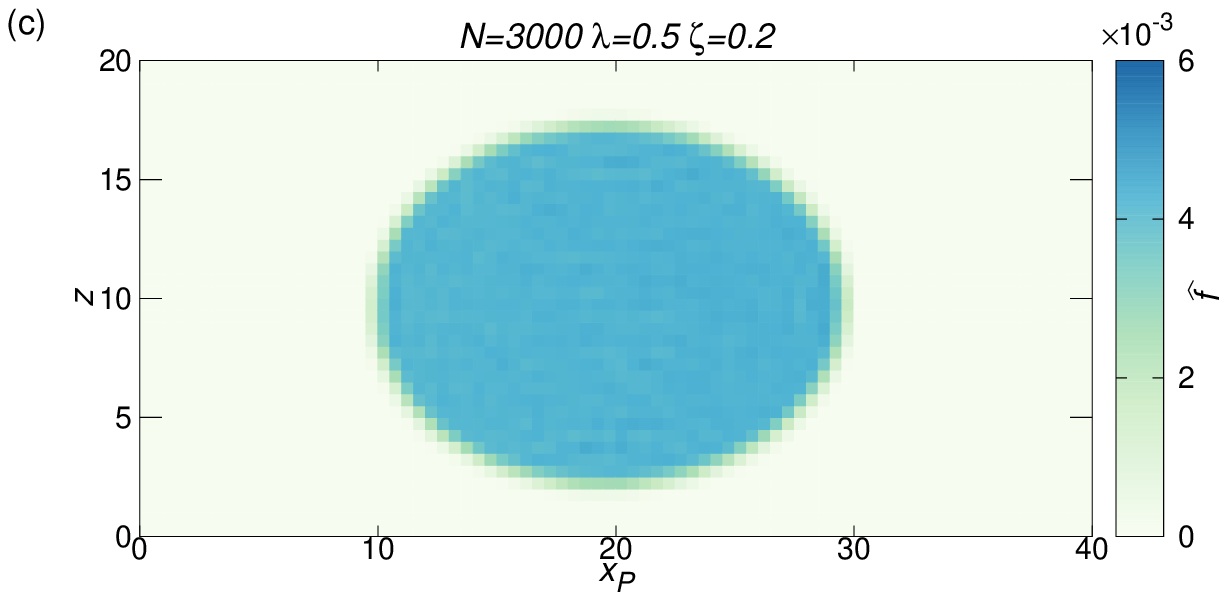}
\vspace{-10mm}\\
\caption{The normalized density distributions $\widehat{f}(\bm{\rho})=\widehat{f}(x_P, z)$
on a vertical cross section of the cluster, where $x_P$ is the coordinate along the moving direction.
A time average is taken over $t\in[1500,21500]$.
The origin is shifted for visibility.
The parameters are 
(a) $\lambda=9.0$, $\zeta=0.2$ and (b) $\lambda=9.0$, $\zeta=2.0$,
and (c) $\lambda=0.5$, $\zeta=0.2$, all for $N_u=3$.
((a) and (b) correspond to the snapshots in 
Fig.~\ref{snapshot}(f) and (g), respectively). 
}
\label{distribution}
\end{figure}

For polarized schools, we also measured
the spatial distributions of the agents $f(\bm{\rho}, t)$
on a vertical cross section of the cluster.
Here, $\bm{\rho}=(x_P,z)$ is the position vector
in the plane containing $\bm{P}(t)$ and the vertical axis.
See Appendix B for details of the definition of $f(\bm{\rho},t)$.
We plot the time-averaged and normalized distribution $\widehat{f}(\bm{\rho})$
in Fig.~\ref{distribution}.
For the standard parameter set ($N_u=3$, $\lambda=9.0$), 
the cluster is elongated vertically and then becomes wing-shaped 
as $\zeta$ increases; see Fig.~\ref{distribution}(a) and (b).
Moreover, the density is larger in the frontal part of the cluster in these cases.
On the other hand, when the attractive interaction is weak ($\lambda=0.5$),
we obtained a horizontally elongated cluster with uniform density 
(Fig~\ref{distribution}(c)).


\section{Results: diffusion in the vertical direction}
\begin{figure}[!t]
\centering
\includegraphics[width=\linewidth,bb=0 0 360 252]{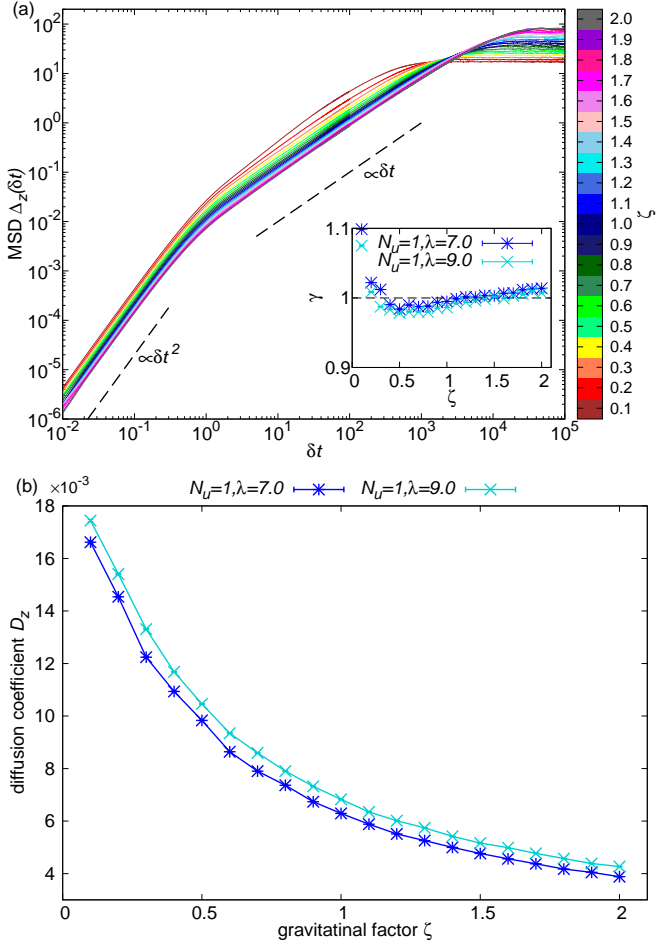}
\caption{MSD in the $z$ direction with $N_u=1,\lambda=7.0,9.0$.
(a) In the case of $\lambda=9.0$, $\Delta_z(\delta t)$ is measured 500 times ($t_0=2500+10n,~n=0,1,\cdots,499$ in Eq. (\ref{DDt})) and averaged.
The inset represents $\zeta$-dependence of the exponent $\gamma$ in Eq. (\ref{diffusion coe}) and the dashed line corresponds to the exponent of normal diffusion $\gamma=1.0$.
(b) The diffusion coefficient $D_z$ as a function $\zeta$.
$\gamma$ and $D_z$ are obtained by fitting the MSD with Eq. (\ref{diffusion coe}), and fitting error is represented by the error bars.
The fitting range is $\delta t\in[5,50]$.
}
\label{MSD}
\end{figure}

To clarify the effect of gravity
on the individual-level motion
in a tornado cluster, 
we study the mean square displacement (MSD) in the $z$ direction:
\begin{equation}
\label{DDt}
\Delta_z(\delta t)=\frac{1}{N}\sum_{i=1}^N\left(c_{z,i}(t_0+\delta t)-c_{z,i}(t_0)\right)^2,
\end{equation}
where $\delta t>0$.
Fig.~\ref{MSD}(a) shows that the MSD increases in proportion to $\delta t^2$ until $\delta t\sim1$, 
and then increases in proportion to $\delta t$ until $\delta t=10^3\sim10^4$, 
and finally takes a constant value.
In other words, an agent performs a ballistic
motion on short timescale $\delta t\sim\tau_0(=1)$ 
and Brownian motion on medium timescale $\delta t\lesssim10^3\sim10^4$, 
before reaching the end of the cluster.

\begin{figure}[!t]
\centering
\includegraphics[width=\linewidth,bb=0 10 360 252]{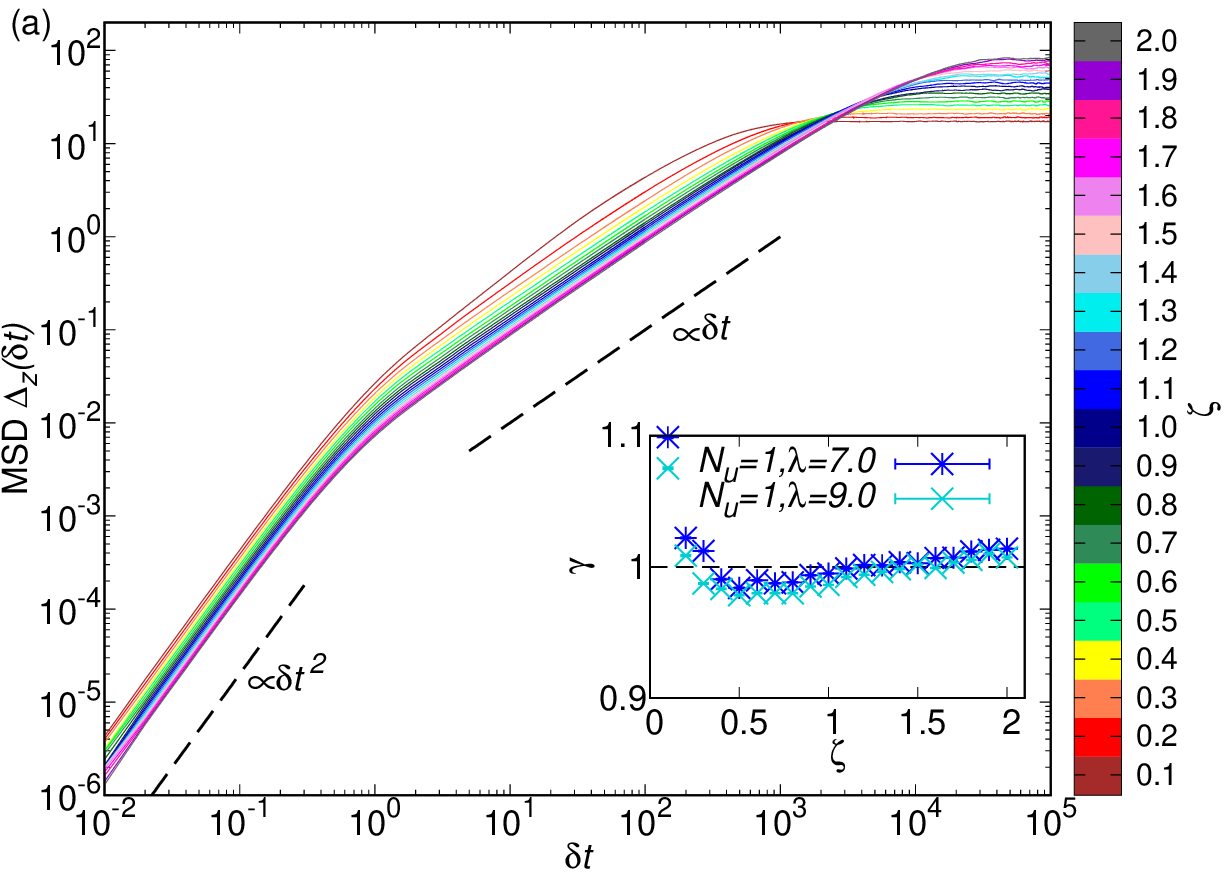}
\end{figure}

We define the vertical diffusion coefficient 
$D_z$ and the exponent $\gamma$ 
to quantify the Brownian motion:
\begin{equation}
\label{diffusion coe}
\Delta_z(\delta t)=2D_z\delta t^\gamma.
\end{equation}
As shown in the inset of Fig.~\ref{MSD}(a), agents perform 
almost normal diffusion ($\gamma=1$)  over a wide range of $\zeta$, 
except the weak superdiffusion ($\gamma\simeq 1.1$) for $\zeta=0.1$.
For $\zeta\gtrsim0.5$, $\gamma$ 
slightly and monotonically increases with $\zeta$.
This is related with a positive correlation between $\gamma$ and the surface area $A\simeq2\pi\overline{R_o}(\overline{R_o}+2\overline{S_3})$ (data not shown.).
The diffusion coefficient $D_z$ decreases as $\zeta$ increases as seen in Fig.~\ref{MSD}(c). 
In addition, $D_z$ increases with the strength of attraction $\lambda$.

\begin{figure}[!b]
\centering
\includegraphics[width=\linewidth,bb=0 0 360 252]{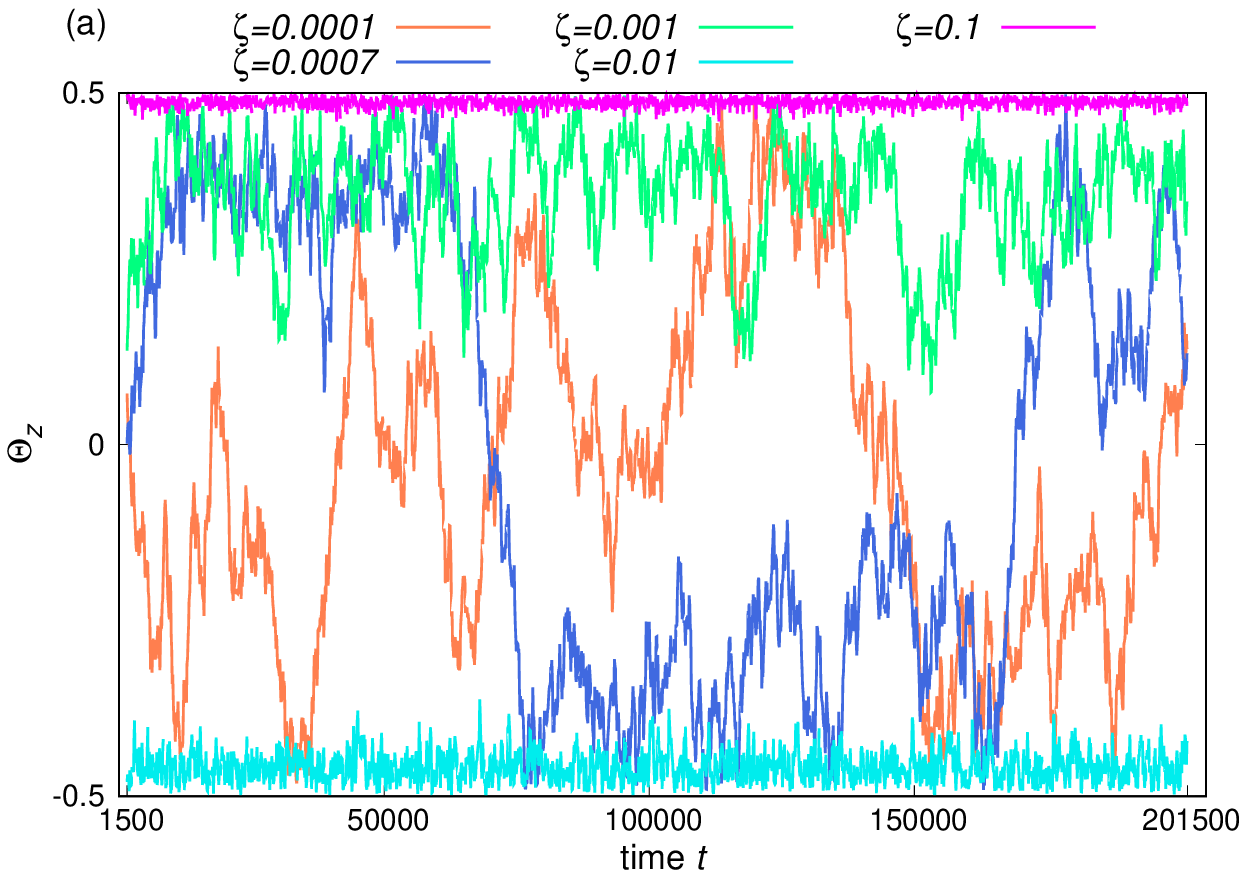}
\end{figure}
\begin{figure}[!b]
\centering
\includegraphics[width=\linewidth,bb=0 0 360 252]{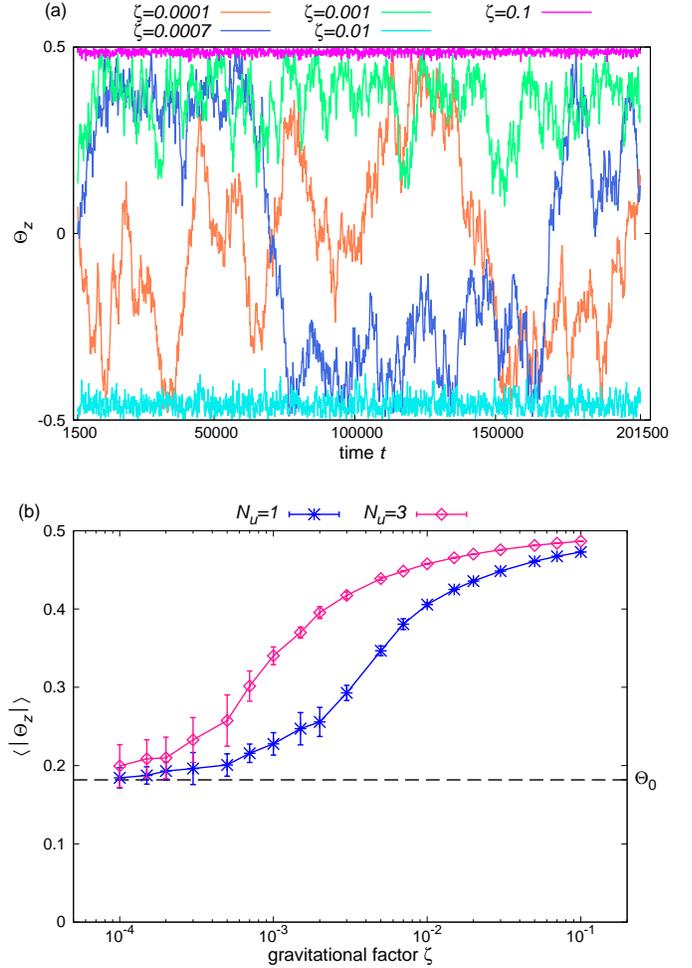}
\caption{$\zeta$-dependence of the motion of the vortex axis.
(a) The typical time evolutions of $\Theta_z(t)$ with $N_u=3,\lambda=7.0$.
(b) $\langle|\Theta_z|\rangle$ as a function $\zeta$ with $N_u=1,3,\lambda=7.0$.
The error bars represent the standard deviation in 10 simulations.
The dashed line represents $\Theta_0$.
}
\label{Thetaz}
\end{figure}

\section{Results: fluctuation of the vortex axis}
We measure the deviation of the vortex axis from the $z$-axis to 
see the effect of gravity on the mobility of the entire cluster.
The angle of the vortex axis and the vertical axis is given by
$\theta_z(t)=\arccos(\bm{M}(t)\cdot\bm{e}_z)\in[0,\pi]$ 
, which we transform to the normalized angle
\begin{equation}
\label{Thetaz def}
\Theta_z(t)=\frac{\theta_z(t)}{\pi}-\frac{1}{2} \in [-0.5,0.5].
\end{equation}
As shown in Fig.~\ref{Thetaz}(a), 
$\Theta_z(t)$ 
shows random motion on a long timescale for $\zeta=0.0001$, 
while it is almost converged to $\pm0.5$ for $\zeta \ge 0.01$.
For the intermediate value $\zeta=0.0007$,
we find flip-flops of the vortex axis 
between $\Theta_z \simeq 0.5$ and $-0.5$.

In order to distinguish between the random motion and the flip-flop, we measure long time-averaged $\langle|\Theta_z(t)|\rangle$, instead of $\langle\Theta_z(t)\rangle$, to quantify the mobility of the vortex axis, where $\langle\cdot\rangle$ 
means time-average over the interval $t\in[1500,201500]$.
Fig.~\ref{Thetaz}(b) shows that $\langle|\Theta_z(t)|\rangle$ 
{increases with $\zeta$, from $\Theta_0 = \frac12 - \frac1\pi \simeq 1.82$ (isotropic limit) 
to 0.5 (vertical limit), for both $N_u=1$ and $N_u=3$.
Note that $\langle|\Theta_z(t)|\rangle=\Theta_0$ means complete random motion of the vortex axis. 
{(See Appendix C for the calculation of $\Theta_0$.)}


\section{Discussions}
Now let us discuss some characteristic features of the collective patterns induced by the gravity.
The giant tornado structure that emerges for $N_u=1$ 
is well approximated by a rotating cylinder in terms of the rotational order parameter.
On the other hand, when $N_u=2$ and $3$, an increase in $\zeta$ 
tends to prevent formation of a rotating cluster and instead promotes 
formation of a polarized school.
The tendency that rotating clusters are more easily obtained for smaller $N_u$ 
is explained by larger sensitivity of each agent to its neighbors~\cite{Ito2021},
and is unchanged under the influence of gravity.

The polarized school obtained for strong attraction ($\lambda=9.0$)
has a high frontal density. 
This is in agreement
with the previous result~\cite{Hemelrijk2008}, but the mechanism could be different.
In our model, attractive forces act on the agents near the surface of a cluster~\cite{Ito2021},
but not on those inside the bulk. The attraction slows down the agents in the front,
while the agents behind them keep moving forward and get jammed at the frontal part.
In fact, the density becomes uniform for weak attraction 
($\lambda=0.5$, Fig.~\ref{distribution}(c)).   
Thus the high frontal density in our model is explained by autonomous control of 
attraction, which is not considered in previous models.
In Ref.~\cite{Hemelrijk2008},
the high frontal density is explained by combination of 
attractive forces and a blind angle.
Another interesting result is that the polarized school is elongated 
vertically for strong attraction and horizontally for weak attraction. 
This is explained as follows.
When the attraction is weak, repulsive forces cause spreading
of agents in the moving direction, as in the case without gravity~\cite{Ito2021}.
On the other hand, when the attraction dominates, 
the cluster is compressed horizontally and the agents are pushed away 
from the dense frontal part toward the vertical direction.  
The agents that are displaced vertically are reoriented toward 
the horizontal direction by the gravitational factor, 
which prevents them to return to the original position.
Thus the cluster is more elongated vertically for larger $\zeta$.
The previous work~\cite{Hemelrijk2008} simulated 
a school of population less than 2000, and obtained 
horizontally elongated clusters, which was explained by sideward orientation.
In experiments, most of the schools in tanks or with small populations 
are elongated horizontally~\cite{Pitcher1979,Partridge1980},
while vertically elongated schools are formed by more than several thousand 
anchovies (\textit{Engraulis mordax}) in deep water~\cite{Smith1970}.
We have checked that reducing the number of agents to $N=300$ in our model
does not significantly change the shape anisotropy  (data not shown).
The effective strength of attraction
might depend on species and environment, and could possibly explain 
the different shape anisotropy in the experiments. However,
it is beyond the scope of the present work and is left for future study.

The MSD in a tornado cluster shows 
the typical time evolution of Brownian motion in a finite domain (See Fig.~\ref{MSD}(a)).
The gravity suppresses vertical motion and therefore
the diffusion constant $D_z$ becomes a decreasing function of $\zeta$ (See Fig.~\ref{MSD}(c)).
(The weak superdiffusion found for $\zeta=0.1$ 
is probably due to fluctuation of the vortex axis from the $z$-axis,
 which has a long correlation time and contributes to the vertical MSD.

The vortex axis of the rotating cluster matches the $z$-axis 
for relatively small $\zeta$ (on the order of 0.01).
This indicates that, if the timescale of the 
reaction to gravitational field $\tau_0/\zeta$ is 
sufficiently longer than the characteristic timescale $\tau_0$, 
the entire cluster shows macroscopic effects of the gravitational field, 
resulting in the vertical alignment of the vortex axis
as experimentally observed \cite{Terayama2015}.
A smaller $N_u$ allows an agent to move more flexibly, 
resulting in larger deviation of the vortex axis
from the $z$-axis (See Fig.~\ref{Thetaz}(b)).

Finally, introducing a blind angle into our model might bring about 
rotating clusters in a wider range of $\zeta$, as found in 
previous models~\cite{Couzin2002,Calovi2014}.


\section{Appendix A: the rotational order parameter of a cylindrical cluster}
The rotational order parameter of a cylindrical cluster of 
height $h$, radius $a$ and uniform density is calculated as follows.
We assume that the agents move horizontally 
with the same radial and tangential velocities $v_r$ and $v_\phi$. 
Regarding the position $\bm{c}=r\bm{e}_r + z \bm{e}_z$ 
and the velocity $\bm{v}= v_r \bm{e}_r + v_\phi \bm{e}_\phi$ of the agents 
as field variables, 
where $\bm{e}_r$ and $\bm{e}_\phi$ are the radial and tangential unit vectors, 
respectively,
we obtain  
\begin{equation}
\label{M cylinder}
M=\left|\frac{1}{\pi a^2 h}
\int_{\mathrm{cylinder}} dV
\frac{\bm{c}\times\bm{v}}{|\bm{c}||\bm{v}|}\right|=w(v_r,v_\phi)\Psi\leri{\frac{h}{2a}},
\end{equation}
where
\begin{eqnarray}
\Psi(\beta)&=&\frac{1}{3\beta}\left\{\beta\sqrt{1+\beta^2}+2\ln\leri{\beta+\sqrt{1+\beta^2}}\right.\nonumber\\
&&\left.+\beta^3\left[ \ln\beta-\ln\leri{1+\sqrt{1+\beta^2}} \right] \right\},
\end{eqnarray}
and $w(v_r,v_\phi)=|v_\phi|/\sqrt{v_r^2+v_\phi^2}$.
In the numerical analysis,
we replaced $w(v_r,v_\phi)$ by its average obtained from the simulation and 
$h$ and $a$ by $2\overline{S_3}$ and $\overline{R_o}$, respectively.
We also plotted the case $w(v_r, v_\phi)=1$ (i.e. $v_r=0$) in Fig.~\ref{order size}.

\section{Appendix B: definition of the spatial distribution function}
First we define the moving orthogonal frame $(x_P, y_P, z)$
so that the polar order parameter $\bm{P}(t)$ is contained in the $x_P$-$z$ plane.
The unit vectors $\bm{e}_x^{(P)}(t)$ and $\bm{e}_y^{(P)}(t)$ along the $x_P$- and $y_P$- 
axis are obtained 
by normalization of $(\bm{I}-\bm{e}_z\otimes\bm{e}_z)\cdot\bm{P}(t)$, 
and $\bm{e}_y^{(P)}(t)=\bm{e}_z\times\bm{e}_x^{(P)}(t)$, respectively.
The spatial distribution on the $(x_P, z)$ plane is defined by
\begin{equation}
\label{rho distri}
f(\bm{\rho},t)=\sum_{i=1}^NH\left(\frac{r_e}{2}-|\bm{e}_y^{(P)}(t) \cdot\bm{c}_i(t)|\right)\delta(\bm{\rho}-\bm{\rho}_i(t)),
\end{equation}
where $H(\cdot)$ is the Heaviside step function, 
$\bm{\rho}=(x_P, z)$,
and $\bm{\rho}_i(t)=\{\bm{I}-\bm{e}_y^{(P)}(t)\otimes\bm{e}_y^{(P)}(t)\}\cdot\bm{c}_i(t)$.
Here we took into account
the agents in a slab of thickness $r_e$ around the vertical center plane.


\section{Appendix C: calculation of $\Theta_0$}
When the motion of the vortex axis is completely random, 
the average value of $|\Theta_z|$ is readily calculated using 
the probability distribution of $\cos\theta_z$, 
which is $P(\cos \theta_z)=1/2$:
\begin{eqnarray}
\label{thetaz deriv}
\Theta_0 &\equiv& 
\int_{-1}^{1}d(\cos\theta_z) P(\cos \theta_z)\left|\Theta_z\right|
\nonumber\\
&=& \frac{1}{2}\int_0^{\pi}d\theta_z\left|\frac{\theta_z}{\pi}-\frac{1}{2}\right|\sin\theta_z
\nonumber\\
&=&\frac{1}{2}-\frac{1}{\pi}.
\end{eqnarray}



\end{document}